\documentclass[twocolumn,aps,prl,showpacs,floatfix]{revtex4}
\usepackage[dvips]{graphicx}
\usepackage{amsmath}
\usepackage{color}
\usepackage{amsfonts}
\usepackage{amssymb}

\DeclareGraphicsExtensions{.eps,.png}
\begin{document}
\title{Clustered chimera states in delay coupled oscillator systems}
\author{Gautam C. Sethia}
\email{gautam@ipr.res.in}
\author{Abhijit Sen}
\affiliation{Institute for Plasma Research, Bhat, Gandhinagar 382 428, India}
\author{Fatihcan M. Atay}
\affiliation{Max Planck Institute for Mathematics in the Sciences, Leipzig 04103, Germany}
\begin{abstract}
We investigate \textit{chimera} states in a ring of identical phase oscillators coupled in a time-delayed and spatially non-local fashion. We find novel \textit{clustered chimera} states that have spatially distributed phase coherence separated by incoherence with adjacent coherent regions in anti-phase. The existence of such time-delay induced phase clustering is further supported through solutions of a generalized functional self-consistency equation of the mean field. Our results highlight an additional mechanism for cluster formation that may find wider practical applications.
\end{abstract}
\maketitle

The study of time delay induced modifications in the collective behaviour of systems of coupled nonlinear oscillators is a topic of much current interest both for its fundamental significance from a dynamical systems point of view and for its practical relevance to modeling of various physical, biological and chemical systems. In real life situations time delay is usually associated with finite propagation velocities of information signals, latency times of neuronal excitations, finite reaction times of chemicals etc.~and in collective oscillator studies it is usually modeled through a time delayed coupling function. Many such past model studies on globally and locally coupled oscillator systems have uncovered interesting and sometimes novel time delay induced modifications of the equilibrium, stability and bifurcation properties of their collective states\cite{SW:89} . While global and local (nearest neighbour) coupling models have traditionally received much attention there is now a growing interest in the collective dynamics of models with non-local couplings \cite{Kur:95,CEVB:97,BC:97,KB:02,Kur:03,AH:05}. Non-local coupling can be relevant to a variety of applications such as in the modeling of Josephson junction arrays \cite{PZWO:93}, chemical oscillators \cite{KB:02,Kur:03,SK:04}, neural networks for producing snail shell patterns and ocular dominance stripes \cite{Mur:89book,ECO:86,Swin:80} etc. One of the striking features of non-locally coupled oscillator systems is that they can support an unusual collective state in which the oscillators separate into two groups - one that is synchronized and phase locked and the other desynchronized and incoherent \cite{KB:02}. Such a state of co-existence of coherence and incoherence does not occur in either globally or locally coupled systems and has been named as a \textit{chimera} state by Abrams \& Strogatz \cite{AS:04}. The nature and properties of this exotic collective state as well as its potential applications are still not fully explored or understood and therefore continue to offer exciting future possibilities. It is not known for example whether such chimera states can exist 
in the presence of time delay in the system and if so then what their characteristics are. This is the principal question we examine in this work through numerical simulations and mathematical analysis of a model system consisting of a ring of densely and uniformly distributed identical phase oscillators that are coupled in a \textit{time-delayed} and spatially non-local fashion. We find that chimera states do indeed exist but acquire an additional spatial modulation such that the single spatially connected phase coherent region of the usual chimera state is now replaced by a number of spatially disconnected regions of coherence with
intervening regions of incoherence. Furthermore the adjacent coherent regions of this clustered chimera state are found to be in anti-phase relation with respect to each other.
To understand the origin and the nature of this pattern we have extended the mean field approach used by Kuramoto \cite{KB:02} and applied it to our system which has a distance-dependent time delay factor in the coupling and have 
derived a functional self-consistency equation. A numerical solution of this self-consistency equation yields a space-dependent order parameter and a space-dependent mean phase function that confirm the existence and explain the nature of the spatial pattern of the oscillator phases.\\

We consider the following model equation representing the continuum limit of a chain of identical phase oscillators arranged on a circular ring $C$,
\begin{eqnarray}
\frac{\partial}{\partial t}\phi(x,t)&&=\omega-\int_{-L}^{L
}G(x-x^{\prime})\nonumber \\
&&\times\sin[  \phi(x,t)-\phi(  x^{\prime},t-\tau_{x,x^{\prime}})  +\alpha]  \,dx^{\prime} \label{field}
\end{eqnarray}
where $2L$ is the system length, $\omega$ is the natural frequency of the oscillator and a closed chain configuration
is ensured by imposing periodic boundary conditions. The kernel $G(x-x^{\prime})$, appropriately normalized to unity over the system length, is taken as,
\begin{equation}
G(x-x^{\prime})=\frac{k}{2(1-e^{-kL})}e^{-k d_{x,x'}} \label{kernel}
\end{equation}  
which provides a non-local coupling among the oscillators over a finite spatial range of the order of $k^{-1}$ which is
taken to be less than the system size. The coupling is time delayed through the argument of the sinusoidal interaction function, namely, the phase difference between two oscillators located at $x$ and $x^{\prime}$ is 
calculated by taking into account the temporal delay for the interaction signal to travel the intervening geodesic (i.e. shortest) distance determined as $d_{x,x'}=min\{|x-x'|,2L-|x-x'|\}$.
The time delay term is therefore taken to be of the form,
$\tau_{x,x^{\prime}}=d_{x,x'}/v$ where $v$ is the signal propagation speed. In the absence of time delay the above equation reduces to the one investigated in \cite{KB:02,Kur:03}. The constant
phase shift term $\alpha$ in the undelayed model breaks the odd symmetry of the sinusoidal coupling function and as discussed in \cite{AS:04,AS:06} it is needed as a tuning parameter for obtaining chimera solutions in the undelayed case. In the presence of time delay however we find that $\alpha$ no longer plays such a critical role since the time delay factor also fulfills a similar function.

\begin{figure}
\includegraphics[scale=0.43]{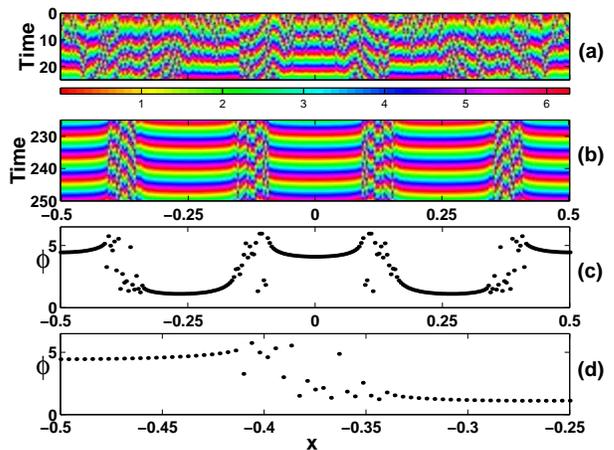}%
\caption{(color online) (a) The space-time plot of the oscillator phases $\phi$ for the parameters $2L=1.0$, $k=4.0$, $1/v=10.24$, $\omega=1.1$ and $\alpha=0.9$ in the early stages of evolution from a random set of initial phases. Panel (b) shows a later time evolution and  panel (c) gives a snapshot of the final stationary state. Panel (d) is a blowup of the region between $x=-0.5$ to $x=-0.25$ giving an enlarged view of an incoherent region and portions of the adjacent coherent regions.}
\label{fig:chimera}%
\end{figure}

We now describe direct numerical simulation results obtained by solving Eq.(\ref{field}) using a large number of discrete oscillators (typically N=256). The set of system parameters chosen for the simulations illustrated here were, $2L=1.0$, $\alpha =0.9$, $k^{-1}=0.25$, $\omega = 1.1$ and $v=0.09765625$ corresponding to a maximum delay time ($\tau_{max}$) in the system of $5.12$.  As discussed in past studies \cite{KB:02,Kur:03}, the choice of appropriate initial conditions is very important for numerically accessing a chimera state. Kuramoto used a random distribution with a Gaussian envelope for the initial distribution of the phases to obtain a chimera solution. For our time-delayed system we find that choosing the initial phases of the oscillators from a uniform random distribution between $0$ and $2\pi$ and then arranging them in a mirror symmetric distribution in space provides a rapid access to a clustered chimera state. Our simulations have been done with the XPPAUT \cite{xpp:02} package using a Runge-Kutta solver (with a small integration time step of $\delta t=0.01$) till a time stationary solution is obtained and tested for independence from discreteness effects by repeating the runs for $N=128$, $256$ and $512$.  In panels (a) and (b) of Fig.\ref{fig:chimera} we show a space time plot of our simulation for the parameters mentioned above in the early stages of evolution (starting from random initial phases) and in the final stages of the 
formation of a {\it clustered chimera} state respectively. Panel (c) shows a snapshot of the spatial distribution of the phases in the final stationary state. We see four coherent regions interspersed by incoherence and also note that the adjacent coherent regions are in anti-phase. Panel (d) is a blowup of the region between $x=-0.5$ to $x=-0.25$ giving an enlarged view of an incoherent region and portions of the adjacent coherent regions. These solutions are
also found to be quite robust and show no signs of instability over arbitrarily large integration times. We have tested
the integrity of the solutions for times well over $100\tau_{max}$.
A detailed parametric study of the stability regions is presently
under progress and will be reported elsewhere.

To gain a better understanding of the nature of this pattern and of the dynamics of its formation we have carried out a mathematical analysis based on the generalized mean field concept as developed by Kuramoto for 
the non-delayed case. For this we first rewrite Eq.(\ref{field}) in terms of a relative phase $\theta(x,t)=\phi(x,t)-\Omega t $ (where $\Omega$ represents a rotating frame in which the dynamics simplifies as much 
as possible such that with the phase-locked portions rotate with this constant drift frequency) as,
\begin{eqnarray}
&&\frac{\partial}{\partial t}\theta(x,t)=\omega-\Omega-\int_{-L}^{L
}G(x-x^{\prime})\nonumber\\
&&\times\sin[\theta(x,t)-\theta(x^{\prime},t-\tau_{x,x^{\prime}})+\alpha+\Omega\tau_{x,x'}]\,dx^{\prime} 
\label{chfield}%
\end{eqnarray}
The key idea behind Kuramoto's analysis of chimera states was the introduction of a mean field like
quantity, namely, a complex order parameter $R e^{i\Theta}$, defined in a manner analogous to what is done for globally coupled systems. For our case we write,
\begin{equation}
R(x,t)e^{i\Theta(x,t)}=\int_{-L}^{L}G(x-x^{\prime})e^{i[
\theta(x^{\prime},t-\tau_{x,x^{\prime}})-\Omega\tau_{x,x^{\prime}}]  }\,dx^{\prime}\label{op}
\end{equation}
The above order parameter differs from the usual definition for global coupling systems in several ways - 
the spatial average of $e^{i\theta}$ is weighted by the coupling kernel $G(x-x^{\prime})$, the phase $\theta$ is 
evaluated in a time delayed fashion and the factor $e^{-i\Omega \tau_{x,x^{\prime}}}$ adds a complex phase to
the kernel $G(x-x^{\prime})$. The latter two features provide a further generalization of Kuramoto's analysis carried out for a non-delayed system \cite{KB:02,Kur:03,AS:04,AS:06}. 

In terms of $R$ and $\Theta$, Eq.(\ref{field}) can be rewritten as :
\begin{equation}
\frac{\partial}{\partial t}\theta(x,t) =\Delta-R(x,t)\sin[\theta(x,t)-\Theta(x,t)+\alpha]\label{geq}
\end{equation}
where $\Delta=\omega-\Omega$. Eq.(\ref{geq}) is in the form of a single phase oscillator equation being driven by
a force term which in this case is the mean field force. To obtain a stationary pattern (in a statistical sense) we require $R$ and $\Theta$ to depend only on space and be independent of time. Under such a circumstance the oscillator population can be divided into two classes: those which are located such that ${R(x)>|\Delta|}$ can approach a fixed
point solution  ($\partial \theta(x,t)/\partial t = 0 $) and the other oscillators that have $R(x)<|\Delta|$ would 
not be able to attain such an equilibrium solution. The oscillators approaching a fixed point in the rotating frame would have phase coherent oscillations at frequency $\Omega$ in the original frame whereas the other set of oscillators would drift around the phase circle and form the incoherent part. Following the prescription provided
by Kuramoto \cite{KB:02,Kur:03} for the undelayed case, we substitute the solutions of Eq.(\ref{geq}) for the two classes of oscillators into the integrand on the right hand side (R.H.S.) of Eq.(\ref{op}) and obtain the following functional self-consistency condition,
\begin{align}
R(x)e^{i\Theta(x)} &  = e^{i\beta}\int_{-L}^{L}G(x-x^{\prime})e^{i[
\Theta(x^{\prime})-\Omega\tau_{x,x^{\prime}}]}\, \nonumber\\
&  \times\dfrac{\Delta-\sqrt{{\Delta}^{2}-R^{2}(x^{\prime})}}{R(x^{\prime})}   \,dx^{\prime}\label{sceq}%
\end{align}
where $\beta=\pi/2-\alpha$. We need to solve for three unknowns --- the functions $R(x)$, $\Theta(x)$ and the quantity $\Delta$. Condition (\ref{sceq}) provides only two equations when we separate its real and imaginary parts. A third condition can be obtained by exploiting the fact that the equation is invariant under any rigid rotation $\Theta(x)\rightarrow\Theta(x)+\Theta_{0}$. We can therefore specify the value of $\Theta(x)$ at any arbitrary chosen point, e.g. $\Theta(L) =0$. We have solved Eq.(\ref{sceq}) numerically by following a three step iterative procedure
consisting of the following steps. We choose arbitrary but well behaved initial guess functions for $R(x)$ and $\Theta(x)$ and use the condition $\Theta(L)=0$ in one of the equations of (\ref{sceq}) to obtain a value for $\Delta$. The initial profiles and the $\Delta$ value so obtained are used to evaluate the R.H.S. of (\ref{sceq}) to generate new profiles for $R$ and $\Theta$. These are next used to generate a new value of $\Delta$ and the 
procedure is repeated until a convergence in the value of $\Delta$ and the functions $R$ and $\Theta$ are obtained.

\begin{figure}
\includegraphics[scale=0.35]{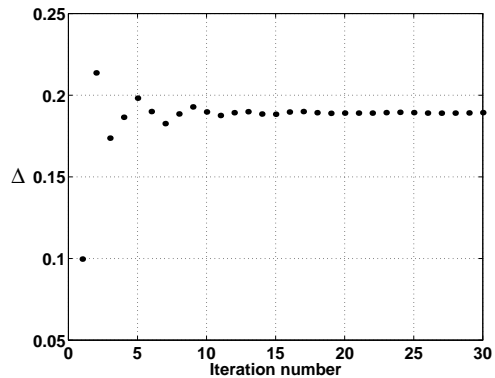}
\caption{Variation of $\Delta$ with the iteration number showing a rapid convergence in the numerical solution of the self-consistency Eq.(\ref{sceq}). The system parameters are identical to those used in the direct solution of Eq.(\ref{field}).} 
\label{fig:delta}
\end{figure}

\begin{figure}
\includegraphics[scale=0.35]{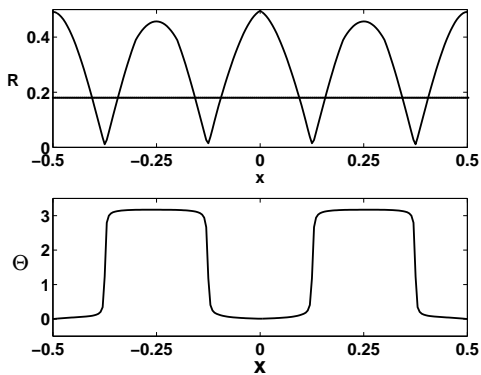}
\caption{Spatial profiles of the amplitude $R$ and the phase $\Theta$ of the order parameter obtained by solving the self-consistency Eq.(\ref{sceq}) by an iterative scheme. The horizontal line in the upper panel, marks the converged value $\Delta=0.189$.}
\label{fig:sc}
\end{figure}

\begin{figure}
\includegraphics[scale=0.43]{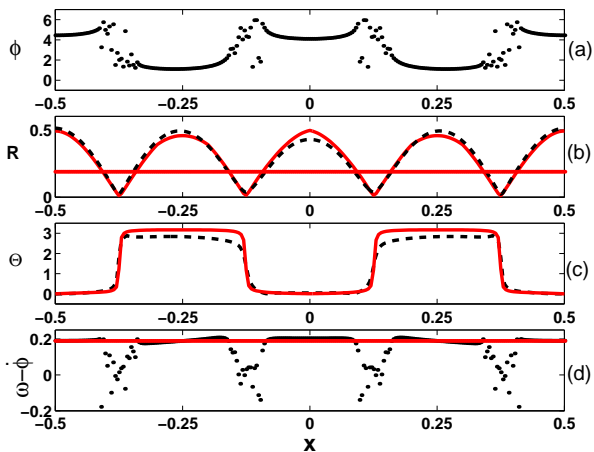}
\caption{(color online) (a) The phase pattern for a clustered chimera state as obtained by direct simulation of Eq.(\ref{field}). The measured spatial profiles of the order parameter ($R$ and $\Theta$) from these simulations are shown in panels (b) and (c) as dashed curves and compared with the solutions from the self-consistency Eq.(\ref{sceq}) shown as solid curves. (d) $\omega - \dot{\phi}$ for the oscillators from a direct simulation of Eq.(\ref{field}). 
The horizontal lines in (b) and (d) mark the converged value of $\Delta=0.189$.}
\label{fig:sc_xpp}
\end{figure}

A MATHEMATICA program incorporating this algorithm was developed and benchmarked against the results for the no-delay case. Fig.\ref{fig:delta} shows the rapid and excellent convergence in $\Delta$ to a unique value of $\Delta=0.189$ for the solution of Eq.(\ref{sceq}) with system parameters chosen identical to the ones that were used to obtain 
a clustered chimera state by a direct solution of Eq.(\ref{field}). The converged spatial profiles of the order parameter ($R$ and $\Theta$) are shown in Fig.\ref{fig:sc} and the converged value of $\Delta$ is marked in the upper panel by the horizontal line. The amplitude of the order parameter ($R$) shows a periodic spatial modulation - peaking at four symmetrically placed spatial locations. The corresponding phases of the order parameter are seen to be in anti-phase for adjacent peaks in $R$. In between the peaks $R$ is seen to dip to very small values at certain locations such that $R(x)<|\Delta|$ which should correspond to the incoherent drifting parts of the chimera. To better appreciate the agreement between the 
direct solutions of Eq.(\ref{field}) and the mean field solutions of Eq.(\ref{sceq}) we have plotted the results together in Fig.\ref{fig:sc_xpp}. As is clearly seen the measured order parameter ($R$ and $\Theta$)  and $\Delta$ from the direct simulations of Eq.(\ref{field}) match well with the results of solving Eq.(\ref{sceq}). The spatial profile of the phases ($\phi$) of the oscillators as obtained from the direct simulation of Eq.(\ref{field}) is shown in the top panel of Fig.\ref{fig:sc_xpp}. We see four coherent regions interspersed by incoherence as expected from the results of solving Eq.(\ref{sceq}). \\
We note from Fig.({\ref{fig:sc}) and Fig.(\ref{fig:sc_xpp}) that both $R$ and $\Theta$ are mirror symmetric (i.e. $R(x)=R(-x), \Theta(x)=\Theta(-x)$), a property that the original phase Eq.(\ref{field}) also possesses. 
Eq.(\ref{field}) is also invariant under the transformation ($\phi(x,t) \rightarrow - \phi(x,t), \omega \rightarrow -\omega, \alpha \rightarrow -\alpha$) and can have solutions with such a symmetry as well, namely, traveling wave solutions given by $\phi(x,t)=\Omega t+\pi qx/L$. In our numerical simulations we find that by changing the initial conditions, but keeping the same system parameters, we can also get traveling wave solutions. There also seems to be a clear correspondence between the number of clusters of the observed chimera state and the wave number $q$ of the co-existent traveling wave solution. For the 4-cluster chimera of Fig.{\ref{fig:sc}} the co-existent traveling wave has $q=2$ and similar results have been obtained for 6-cluster ($q=3$) and 8-cluster ($q=4$) chimera solutions.

To conclude, we have demonstrated for the first time the existence of chimera type solutions in a time-delayed system of non-locally coupled identical phase oscillators. Time delay is found to lead to novel clustered states with a number of spatially disconnected regions of coherence with intervening regions of incoherence. The adjacent coherent regions of this clustered chimera state are found to be in anti-phase relation with respect to each other. Our numerical simulations are further validated and explained through solutions of a generalized functional self-consistency equation of the mean field. The mean field parameters (the amplitude and phase of the complex order parameter) clearly reflect the modulated nature of the effective driving force on each oscillator and lead to the resultant pattern of phase distribution seen in the clustered chimera state. Thus time delay offers an additional mechanism for cluster formations in dynamical systems and model systems incorporating time delay may provide a useful paradigm for studying this phenomenon. Our results can be usefully extended to higher dimensions e.g. to examine the influence of time delay on spiral wave based chimeras in two dimensions \cite{SK:04} and may also help provide insights into experimental observations of clustered states that are generic to many chemical and biological systems\cite{GR:94}. 

GCS thanks P. Abbott and S. Richardson for help on Mathematica. FMA and GCS acknowledge the hospitality of MPI-PKS, Dresden, Germany, at DYONET 2006.



\end{document}